\documentclass[draft,showpacs,preprintnumbers,amsmath,amssymb,fourier]{revtex4-2}
\usepackage{graphicx}
\usepackage{color}

\usepackage{amsmath, amsthm, amsfonts}
\usepackage{amssymb}

\theoremstyle{definition}

\theoremstyle{remark}

\newcommand{\abs}[1]{\left\vert#1\right\vert}

\begin{document}

\title{Linearized Horndeski Theory with a Potential in the Solar System Regime}

\begin{abstract}
In this paper, the weak-field behavior of linearized Horndeski theory is studied, with emphasis on the role of a scalar potential with a nonvanishing minimum. In this regime, the minimum of the potential acts as an effective source of background curvature and produces a contribution similar to a cosmological constant. The analysis is restricted to the linear approximation, where nonlinear screening effects such as the Vainshtein mechanism can be consistently neglected. Within this framework, the consistency of the theory with Solar-System phenomenology in the weak-field limit is examined, and possible deviations from General Relativity depending on the model parameters are discussed. To this end, the linearized field equations for a static point mass are derived, the corresponding geodesic motion is investigated, and the resulting weak-field effects in classical Solar-System observables, including perihelion advance, light deflection, and gravitational redshift are analyzed. The analysis further focuses on the limiting regimes of very light and very heavy scalar fields. In the very light scalar field regime, consistency with Solar System phenomenology requires sufficiently large values of the coupling parameter $\zeta$, thereby suppressing the scalar contribution at local scales and keeping deviations from General Relativity negligible. In the very heavy scalar field regime, the scalar-mediated interaction acquires a short range and becomes dynamically suppressed, leading to weak-field predictions that are practically indistinguishable from those of General Relativity. Nevertheless, geometric terms associated with the minimum of the scalar potential may persist at linear order in the metric perturbations, depending on the value of $\zeta$.
\end{abstract}


\author{Hatice Özer}
\affiliation{Department of Physics, Faculty of  Sciences,  Istanbul University, 34134  Istanbul, Türkiye\\
ORCID:0000-0002-6325-0598}

\maketitle

\newpage

\section{Introduction}
 Einstein’s General Theory of Relativity (GR) has stood for over a century as a foundational framework of modern gravitational physics, consistently validated by a broad range of high-precision experiments and astronomical observations  \cite{Misner,Will1,Will2}. Despite its remarkable success, GR does not fully account for several key cosmological phenomena, including the late-time accelerated expansion of the universe, the nature of dark energy and dark matter, and its incompatibility with quantum mechanics \cite{Sotiriu,Nojiri,Capoziella,Clifton,Turner,Sahni,Padmanabhan,Sahni2,Upadhye}. One of the greatest challenges faced by the General Theory of Relativity is explaining the accelerated expansion of the universe \cite{Perlmutter,Riess,Riess2,Eisenstein,Tonry}. Although this acceleration can be explained by introducing a cosmological constant into Einstein’s equations, the fact that this constant must have an extremely fine-tuned value leads to a serious fine-tuning problem \cite{Weinberg,Adler}. To overcome these issues, scalar–tensor theories have been developed, proposing the idea of a dynamic scalar field that drives the universe’s expansion \cite{Jordan,Brans,Fujii}. Scalar–tensor theories play a crucial role in exploring the fundamental nature of gravity beyond General Relativity. Instead of introducing unknown forms of matter or energy, these theories modify the gravitational interaction itself, thereby providing a unified theoretical basis for explaining both the accelerated expansion of the universe and dark matter–like effects observed on galactic and cosmological scales. Among these, one of the most comprehensive formulations is Horndeski theory \cite{Horndeski}, which stands out as the most general scalar-tensor theory yielding second-order field equations, thereby ensuring theoretical consistency and avoiding Ostrogradsky instabilities \cite{Ostrogradsky}. By introducing derivative couplings between the scalar field and curvature tensors, it provides cosmologically viable models that can account for the universe’s accelerated expansion. As a unifying framework, it includes Brans–Dicke theory, f(R) gravity, and k-essence as specific limits and provides a consistent and observationally testable model for modified gravity theories. The Horndeski theory, while offering both theoretical and observational advantages, must still reproduce the predictions of GR in the appropriate limits. Solar System observations therefore play an important role in asessing the phenomenological viability of alternative theories of gravity in the weak-field regime. So far, all gravitational phenomena observed in the Solar System and in binary systems are consistent with the predictions of GR. Consequently, phenomenologically acceptable Horndeski models should remain compatible with classical Solar-System tests, such as light deflection, perihelion advance and gravitational redshift. The primary aim of this work is to analyze the weak-field predictions of linearized Horndeski theory with a scalar-field potential in the presence of effective de Sitter-like curvature contributions, with particular emphasis on the behavior of the theory at Solar-System scales.

In this work, the weak-field solutions of Horndeski gravity are analyzed in the linearized framework. This theory generalizes Einstein’s General Relativity by introducing a dynamical scalar field governed by a non-trivial potential that can effectively mimic a cosmological constant in the linearized regime. Similar to Brans–Dicke theory with a potential, a nonvanishing minimum may behave like a cosmological constant \cite{Hatice}. However, in the linearized Horndeski theory with a potential this interpretation applies only at the linear level and should not be interpreted as reflecting the full nonlinear theory. Several works have explored weak-field solutions and their astrophysical predictions in different modified gravity models \cite{Pechlaner,Wagoner,Stelle,Will3,Steinhardt,Barros,Olmo1,Olmo2,Olmo3,Perivolaropoulos,Berry,Berry2,Alsing,Hohmann,Hohmann2,Hatice2, Setare}. Similar Solar-System and parametrized post-Newtonian analyses have also been carried out in Horndeski gravity and related scalar-tensor theories \cite{Hou,Gonzalez,DellaMonica,Sakstein}. In most of these analyses, asymptotic flatness is adopted as a standard assumption. Nevertheless, alternative formulations involving curved background geometries and nontrivial configurations in Horndeski gravity have also been considered in the literature \cite{Bernabeu2011,Babichev2014,Cisterna2015,Jha2023,Nunes2015,Santos}. Motivated by these considerations, the weak-field solutions of Horndeski theory are investigated here within a linearized weak-field regime containing effective de Sitter-like curvature corrections. Within this framework, the scalar potential plays a central role in determining both the local weak-field behavior and the large-scale geometric properties of the theory. In particular, a nearly constant or slowly varying potential can give rise to an effective de Sitter-like behavior at linear order, thereby providing a possible explanation for the observed accelerated expansion of the universe. In the present analysis, however, this effective de Sitter-like contribution is treated only within the local weak-field approximation. More precisely, the terms proportional to the potential minimum are retained as leading-order curvature contributions in the perturbative expansion, while higher-order mixed terms involving both the background curvature scale and metric perturbations are neglected consistently. Consequently, the effective de Sitter-like background considered here should be understood as a local perturbative curvature correction valid within the adopted weak-field regime rather than as an exact nonlinear cosmological background solution. Nevertheless, the scalar potential remains physically important since it governs both the effective curvature scale and the scalar-field mass appearing in the weak-field dynamics. It has been suggested that a scalar field with a slowly varying potential could be responsible for both inflation \cite{Linde} and dark energy \cite{Copeland}. Moreover, the inclusion of a potential term in scalar–tensor theories endows the scalar field with an effective mass, which restricts its interaction range and can lead to a suppression of deviations from General Relativity in the weak-field regime \cite{Will2,Hatice2,Alsing}. On the other hand, a massless scalar field mediates long-range interactions, potentially leading to observable deviations from General Relativity. Therefore, the scalar potential plays a crucial role in determining the local behavior of the theory and in constraining its phenomenological implications at Solar-System scales. Such extensions must remain compatible with the well-established predictions of General Relativity in the Solar System. Indeed, GR has been remarkably successful in explaining classical tests such as light bending, perihelion shift and gravitational redshift \cite{Eddington,Shapiro,Diche,Berti,Will1,Will3}, and any viable alternative theory is expected to remain compatible with these results in the appropriate weak-field regime. Since potential terms can induce small but measurable scalar-mediated deviations from GR, we investigate how such effects arise in the weak-field regime and analyze the resulting behavior of the theory within this framework. Although Horndeski theory has received extensive attention particularly in the context of weak-field expansions, post-Newtonian parameters \cite{Capoziella,Clifton,Sotiriu,Kobayashi,Felice,Minamitsuji,Babichev,Silva,Lima,Heisenberg}, many existing studies either neglect the scalar potential or assume asymptotically flat backgrounds. Classical tests such as light deflection and the perihelion precession have been extensively studied in the context of general scalar-tensor theories, however detailed analyses within Horndeski models that include a scalar potential are still limited. Accordingly, we study the weak-field behavior of these models within the linearized framework, where nonlinear screening mechanisms such as the Vainshtein effect are not operative, and examine how the presence of a scalar potential influences the resulting behavior.

In this study, the weak-field solutions of Horndeski theory are obtained in the presence of effective de Sitter-like curvature contributions arising at linear order. This makes it possible to investigate how the minimum of the scalar-field potential modifies the local space-time structure and weak-field gravitational dynamics. By choosing an appropriate gauge, the linearized field equations are derived and solved for a static point-like source. The obtained solutions are then transformed into isotropic and Schwarzschild-like coordinate systems in order to enable comparison with classical Solar-System observations in the weak-field regime. The physical properties of these solutions are analyzed by examining the geodesic structure in Schwarzschild-type coordinates. Throughout this work the analys is restricted to the linearized regime of Horndeski theory, where nonlinear screening mechanisms such as the Vainshtein effect are not taken into account. The aim of this work is not to provide a fully nonlinear description of the theory, but rather to focus on its linearized behavior. Within this framework, the influence of a scalar potential on the weak-field dynamics is investigated. In particular, classical observables such as perihelion shift, light deflection, and gravitational redshift are analyzed within the linearized approximation, with emphasis on the role of the source mass and the scalar potential, and with the analysis restricted to the weak-field regime.In addition, the Cassini result on the effective PPN parameter $\gamma$ is used here only as an estimate of the observational sensitivity of Solar-System measurements within the adopted linearized weak-field framework, rather than as a direct observational constraint on the full nonlinear Horndeski theory. Since strong-field effects are not studied, it is convenient to use the linearized version of the field equations of this theory. This choice allows us to investigate the generic features of Horndeski theory in this regime without specifying the exact forms of the scalar functions $G_i(\phi)$ and $K(\phi)$. To remain at the linearized level, we neglect second-order terms such as $K(0) h_{\mu\nu}$, $K(0) \varphi$ and $h_{\mu\nu} \varphi$.

This paper is organized as follows. In  section II, the linearized field equations of the  Horndeski theory in the presence of a scalar potential are investigated in a background geometry that becomes asymptotically non-flat due to the minimum of the potential. In section III, the case of a static point-mass source is examined in the massive Horndeski theory with an arbitrary potential. The field equations are solved in the adopted gauge and the resulting solutions are transformed into isotropic and Schwarzschild-type coordinates. The behavior of the theory in the linearized weak-field regime relevant to Solar-System scales is examined in Sections IV–VI: the advance of perihelion in Section IV, the deflection of light in Section V, and the gravitational redshift in Section VI. The paper ends with a brief discussion.

\section{Weak Field Equations}
The action of Horndeski theory is given by \cite{Horndeski}
\begin{equation}
	S= \int d^4x \sqrt{-g}(L_2+L_3+L_4+L_5)+\int dx^4 \sqrt{-g} L_m,
\end{equation}
where Horndeski lagrangian densities are,
\begin{align}
	L_2&= K(\phi,X), \quad L_3= -G_3(\phi, X) \Box\phi, \nonumber \\
	L_4&= G_4(\phi,X) R + G_{4,X} \left[(\Box\phi)^2 - (\nabla_\mu\nabla_\nu\phi)(\nabla^\mu\nabla^\nu\phi)\right], \label{Hornactions} \\
	L_5&= G_5(\phi,X) G_{\mu\nu} \nabla^\mu\nabla^\nu\phi - \frac{1}{6}G_{5,X} \left[(\Box\phi)^3 
	- 3(\Box\phi)(\nabla_\mu\nabla_\nu\phi)(\nabla^\mu\nabla^\nu\phi) 
	+ 2(\nabla^\mu\nabla_\alpha\phi) (\nabla^\alpha\nabla_\beta\phi) (\nabla^\beta\nabla_\mu\phi) \right]. \nonumber
\end{align}  
Here $g$ is the determinant of the metric tensor, $\Box=g_{\mu\nu}\nabla^\mu\nabla^\nu$ is the D'Alembertian operator, 
the functions $K$, $G_3$,  $G_4$,$G_5$ are arbitrary functions of $\phi$ and $X$, where   $X=-\nabla_\mu\phi\nabla^\mu\phi/2$ is a canonical kinetic term. In this work, the motion of test particles is investigated in the Jordan frame, where the scalar field is part of the gravitational interaction and is non-minimally coupled to the curvature. In this frame, test particles follow geodesics determined by the Jordan-frame metric, meaning that their trajectories are governed solely by the spacetime geometry and do not involve a direct coupling to the scalar field. Accordingly, observable quantities such as the deflection angle of light and orbital precession can be computed using the metric tensor within the linearized framework and examined in terms of their distinctive behavior in the weak-field regime. On the other hand, for a broad class of Horndeski theories, the action (1) can be reformulated in the Einstein frame, where matter fields are no longer minimally coupled to the metric but instead interact through a geometry that depends on the scalar field. As a result, test particles no longer follow the geodesics of the metric and are subject to an additional force, often referred to as the fifth force. The analysis of orbital trajectories and light bending in this frame requires the inclusion of the extra force contributions induced by the scalar field, thereby making the calculations more difficult and the physical interpretation less straightforward. Moreover, transitioning to the Einstein frame in Horndeski theory is significantly more complicated than in standard scalar-tensor theories. This is due to the fact that in simpler theories the necessary rescaling of the metric can be achieved through a conformal transformation based on a scalar function. Nevertheless, in Horndeski theory the function $G_4$ also depends on X, which involves the derivatives of the scalar field $\phi$. It is often necessary to perform a disformal transformation of the form \cite{Will2}
\begin{equation}
	\tilde{g_{\mu\nu}}=\Omega^2(\phi,X)g_{\mu\nu}+W(\phi,X)\partial_{\mu}\phi\partial_{\nu}\phi
\end{equation}
where $\Omega^2(\phi,X)$ is the conformal factor and $W(\phi,X)$ is the disformal factor, both being suitably chosen functions of  $\phi$ and X. Disformal transformations introduce an anisotropic stretching that depends on the direction of $\partial_{\mu}$ and they can alter the causal structure of spacetime. While conformal transformations primarily corresponds to simple rescalings of the metric, disformal transformations induce anisotropic and derivative-dependent modifications to the geometry, substantially complicating the mathematical and physical analysis of the transformed theory. For this reason, bringing the Horndeski action (1) into the Einstein frame is  generally far less straightforward than in other scalar-tensor theories. In the Jordan frame, although the scalar field does not give rise to a direct fifth force, it can influence the dynamics indirectly through its coupling to the metric. In this study, we  restrict our analysis to the linearized regime, where such effects remain perturbative. For the reasons outlined above, we consider the Jordan frame as an appropriate framework for studying the weak-field behavior of the theory at Solar System scales. Since the gravitational field is weak in the regime considered here, we employ the weak-field approximation and expand the metric and scalar field accordingly, 
\begin{align}
	g_{\mu\nu} &= \eta_{\mu\nu} + h_{\mu\nu},\quad  g^{\mu\nu}=\eta^{\mu\nu}-h^{\mu\nu}, \\
	\phi &= \phi_0 + \varphi,
\end{align}
where $\eta_{\mu\nu}$ is the Minkowski metric and $\phi_0$ is a constant background value of the scalar field. The quantities $h_{\mu\nu}$ and $\varphi$ denote the metric and scalar perturbations, respectively, and satisfy $|h_{\mu\nu}| \ll 1$ and $|\varphi| \ll 1$. Within the present weak-field expansion, the field equations are linearized by retaining only terms up to first order in the perturbative quantities. Consequently, products of perturbations are consistently neglected as higher-order contributions. This includes not only explicitly quadratic terms such as $h_{\mu\nu}\varphi$, but also mixed perturbative terms appearing in the expansion procedure.

Let us note that the function $K$ appearing in Eq.(1) contains both the kinetic and potential contributions of the theory. In this work, we investigate Solar-System effects in the presence of a non-flat background geometry, which may originate from a potential contribution contained in the function $K$. Therefore, the general case $K\neq0$ is considered. In the present analysis, the nonvanishing minimum of the scalar potential is not interpreted as defining an exact nonlinear de Sitter background solution. Instead, within the adopted weak-field approximation, the potential minimum contributes only through local perturbative curvature corrections proportional to $K(0)r^2$. Accordingly, the analysis is restricted to the exterior weak-field region satisfying
\[
r>R,
\qquad
\frac{|K(0)|r^2}{G_4(0)}\ll1,
\]
where $R$ denotes the radius of the gravitating body. These conditions ensure that the contribution proportional to $K(0)r^2$ remains a perturbatively small correction to the metric.
Since a nonvanishing minimum of the scalar potential generates an effective background curvature, asymptotic flatness is not expected for $K(0)\neq0$. Therefore, the present solution is not intended to describe the asymptotic region $r\rightarrow\infty$. Asymptotic flatness is recovered only in the limit $K(0)\rightarrow0$, where the background curvature vanishes. For the Solar-System scales considered in this work, the condition
\[
\frac{|K(0)|r^2}{G_4(0)}\ll1
\]
is well satisfied, ensuring the self-consistency of the adopted perturbative treatment. In this regime, the spacetime may be approximated locally by Minkowski geometry. Furthermore, the curvature effects remain perturbatively small, so that the terms proportional to $K(0)$ are consistently retained as the leading-order curvature corrections. In particular, for the Solar-System scales considered in this work, the quantity $|K(0)|r^2/G_4(0)$ remains extremely small. Within the adopted perturbative expansion, the terms proportional to $K(0)$ are retained as leading-order curvature contributions associated with the effective local de Sitter-like correction. However, mixed terms such as $K(0)h_{\mu\nu}$ contain both the small background-curvature parameter and the perturbative metric fluctuations simultaneously, and therefore represent higher-order quantities within the perturbative ordering adopted in the present analysis. Such terms are consequently neglected consistently in the linearized approximation. Accordingly, the effective de Sitter-like contribution appearing in the weak-field solution should not be interpreted as an exact global background geometry, but rather as a local perturbative curvature correction valid within the adopted weak-field regime. In order to derive the linearized field equations consistently within the weak-field approximation, the function $K(\phi,X)$ must also be linearized appropriately. Since the kinetic term
$
X=-\frac{1}{2}\partial_\mu \phi \,\partial^\mu \phi$ 
is quadratic in the scalar perturbation, it is neglected at linear order, and thus the function $K(\phi,X)$ effectively reduces to its potential part, denoted by $K(\phi)$. In alternative theories of gravity, the choice of scalar potential plays a crucial role in determining the physical predictions and cosmological behavior of the theory. Different forms of the potential modify the mass, dynamics, and interaction range of the scalar field. The background scalar field $\phi_0$ is assumed to correspond to a stable vacuum configuration of the theory, namely a minimum of the effective potential. This assumption ensures the stability of the scalar field under small perturbations and allows for a consistent weak-field expansion around a static background configuration relevant for Solar-System analyses. As a result, small perturbations of the scalar field oscillate around the background value instead of becoming unstable, implying the condition $K_{,\phi}(0)=0$. Under this assumption, the leading contribution governing the dynamics of the scalar perturbation arises from the second derivative of the potential, which determines the effective mass and interaction range of the scalar degree of freedom in the weak-field regime. Expanding the potential around the background scalar field value $\phi=\phi_0$e obtain
\begin{eqnarray}\label{Kexpansion}
	K(\phi)&=&K(0)+K_{,\phi}(0)\varphi+\frac{1}{2}K_{,\phi\phi}(0)\varphi^2+\ldots,\\
	K_{,\phi}(\phi)&=&K_{,\phi}(0)+K_{,\phi\phi}(0)\varphi+\ldots,\nonumber
\end{eqnarray}
where, for notational simplicity, we use the shorthand notation $
K(0)\equiv K(\phi_0,0)$. The constant part of the potential, $K(0)$, is nevertheless retained since it acts as an effective background source term generating the de Sitter-like curvature contribution to the spacetime geometry at linear order. In the adopted weak-field expansion, the quantity $K(0)$ is treated as a small background-curvature parameter whose leading contribution is proportional to $K(0)r^2$. Accordingly, while the background curvature contribution sourced by $K(0)$ is preserved at leading order, mixed terms such as $K(0)h_{\mu\nu}$ and $K(0)\varphi$ contain both the small background-curvature scale and perturbative field fluctuations simultaneously, and are therefore treated as higher-order quantities within the perturbative expansion. Such terms are consequently neglected consistently in the linearized approximation. Similar perturbative treatments have also been employed in weak-field analyses of GR in the presence of cosmological background contributions \cite{Bernabeu,Bernabeu2011}. Hence, the weak field equations, up to the linear order in $h_{\mu\nu}$ and $\phi$ become,
\begin{align}
	&-\frac{1}{2} K(0)\eta_{\mu\nu} + G_4(0)G^{(1)}_{\mu\nu}(0)-G_{4,\phi}(0)(\partial_\mu\partial_\nu\varphi-\eta_{\mu\nu}\Box\varphi) = 0, \\
	&K_{,\phi}(0) + [K_{,X}(0)-2G_{3,\phi}(0)]\Box\varphi + K_{,\phi\phi}(0)\varphi + G_{4,\phi}(0)R^{(1)} = 0, \label{leq2}
\end{align}
where $G^{(1)}_{\mu\nu}$ is the linearized Einstein tensor and  $R^{(1)}$ is the linearized Ricci scalar. Here we use the notation $K_{,X}=\frac{\partial K}{\partial X}$ and $K_{,\phi}=\frac{\partial K}{\partial \phi}$ for a function K of  $\phi$ and $X$. It is important to emphasize that all differential operators are defined with respect to the flat (Minkowski) background spacetime. The weak field equations of Horndeski theory in a non-flat background geometry  are obtained under the conditions
\begin{equation}
	K(0)\neq 0,\quad  K_{,\phi}(0)=0.
\end{equation} 
This form of the potential allows the scalar field to oscillate around a minimum and remain in a stable configuration, which leads the scalar field to acquire a mass and exhibit short-range effects. Moreover, the minimum of the scalar-field potential gives rise to an effective curvature contribution that behaves as a de Sitter-like correction at linear order within the weak-field expansion. The quadratic term in the expansion, on the other hand, represents a dynamical contribution that drives the scalar field toward the minimum at $\phi_0$. This structure therefore provides a dynamically generated cosmological-constant-like contribution within the adopted perturbative regime. In this context, the aim of the present work is to investigate how such a potential structure modifies the weak-field behavior of linearized Horndeski theory and its Solar-System predictions.
To simplify the tensor field equation (7), we introduce the auxiliary tensor field as 
\begin{equation}
	h_{\mu\nu}= \theta_{\mu\nu} - \frac{1}{2}\eta_{\mu\nu}\theta-\eta_{\mu\nu}\sigma\varphi,
\end{equation}
where $\sigma=\frac{G_4,\phi(0)}{G_4(0)}$ and impose the Lorentz gauge condition $\partial^\mu \theta_{\mu\nu}=0$. 
The field equations of linearized Horndeski theory become
\begin{equation}
	\Box\theta_{\mu\nu}=-\frac{16\pi T_{\mu\nu}}{G_4(0)}-\frac{K(0)\eta_{\mu\nu}}{G_4(0)} 
\end{equation}
\begin{equation}
	(\Box-m_s^2)\varphi= 8\pi T\chi+2K(0)\chi
\end{equation}
where
\begin{equation}
	\chi=\frac{G_{4,\phi}(0)}{G_4(0)[K_{,X}(0)-2G_{3,\phi}(0)+3G^2_{4,\phi}(0)/G_4(0)]}
\end{equation}
The scalar field is massive, with the corresponding scalar mass defined as
\begin{equation}\label{ms}
	m_s^2=- \frac{ K_{,\phi\phi}(0)}{K_{,X}(0)-2G_{3,\phi}(0)+\frac{3G^2_{4,\phi}(0)}{G_4(0)}},
\end{equation}
 For physical consistency, the analysis is restricted to the parameter region satisfying
$m_s^2>0$. This condition guarantees a stable Yukawa-type suppression of the
scalar field and avoids tachyonic instabilities. Physically, this ensures that
small scalar perturbations remain stable around the background vacuum
configuration instead of exhibiting exponentially growing behavior.
From Eq. (14), the requirement $m_s^2>0$ implies
\begin{equation}
	\frac{K_{,\phi\phi}(0)}
	{K_{,X}(0)-2G_{3,\phi}(0)+\dfrac{3G_{4,\phi}^2(0)}{G_4(0)}}
	<0.
\end{equation}
This condition follows directly from the positivity of the scalar mass squared and defines the parameter region for which the scalar field remains free of tachyonic instabilities. Throughout the present analysis, it is further assumed that the denominator is nonzero, so that the effective scalar mass is well defined.

As clearly indicated by Eq.(11) the structure of the tensor equation closely resembles that of $GR\Lambda$ and $BD\Lambda$ and BDV field equations \cite{Hatice2}. A comparison with the corresponding equations in Lorentz gauge \cite{Bernabeu} shows that K(0) must be negative and approach $K(0)\rightarrow -2\Lambda\, G_4(0) $ to reproduce the GR$\Lambda$ 
 de Sitter limit. Considering the scalar field equation, Eq.(14) indicates that a scalar potential with a nonzero second derivative inevitably generates an effective mass term for the Horndeski scalar field. This behavior is typical in Yukawa-type corrections to Newtonian gravity and in the mass term of the Klein–Gordon equation. In alternative theories of gravity, the mass of the scalar field is determined by the structure of the scalar part of the theory and the model parameters. This mass governs the propagation of scalar perturbations through the Klein–Gordon–type equation and therefore determines how the field behaves throughout spacetime. As a result, the scalar mass plays an important role in the theoretical description of the system, as discussed in the following sections.
\section{Solutions to the Linearized Field Equations for a point mass}
In this section, we consider a static point mass solution as a source for massive Horndeski theory. We assume a point particle positioned at $\bar{r}=0$ where $\bar{r}^2=\bar{x^2}+\bar{y^2}+\bar{z^2}$ and the energy momentum tensor of the point particle is described by 
\begin{equation}
	T_{\mu\nu}=m\delta(\bar{r})diag(1,0,0,0)
\end{equation}
Inserting the trace of the energy-momentum tensor into Eq.(12) and assuming a static case, we obtain the scalar field solution as follows,
\begin{equation}
	\varphi(\bar{r})=\frac{2m\chi}{\bar{r}}e^{-m_s\bar{r}}+\frac{K(0)\chi}{3}\bar{r}^2
\end{equation}
After performing a suitable coordinate transformation to bring the solution into static form, the resulting metric becomes
\begin{equation}
	ds^2=-\Big[1-\frac{2m}{G_4(0)\bar{r}}+\frac{K(0)}{6G_4(0)}\bar{r}^2-\sigma\bar{\varphi}\Big]\bar{dt^2}+
{\sum_{i=1}^3}
	\left[1+\frac{2m}{G_4(0)\bar{r}}+\frac{K(0)}{12G_4(0)}(\bar{r}^2+3\bar{x}_i^2)-\sigma\bar{\varphi}\right]d\bar{x_i}^2,
\end{equation}
To express this solution in isotropic coordinates, we apply the following coordinate transformations,
\begin{equation}
	\bar{x}^i=x'^i-\frac{K(0)}{24G_4(0)
	}{x'^{i}}^3	\nonumber
\end{equation}
\begin{equation}	
	\bar{t}=t' \\,
\end{equation}
Under these transformations, the metrik perturbation terms take the following form
\begin{equation}
	h'_{00}=\frac{2m}{G_4(0)r'}-\frac{K(0)}{6G_4(0)}r'^2+\sigma\varphi'
\end{equation}
\begin{equation}
	h'_{ij}=\Big(\frac{2m}{G_4(0)r'}+\frac{K(0)}{12G_4(0)}r'^2-\sigma\varphi'\Big)\delta_{ij}
\end{equation}
\begin{equation}
	\varphi'({r})=\frac{2m\chi}{r'}e^{-m_sr'}+\frac{K(0)\chi r'^2}{3}
\end{equation}
It is apparent that the contributions of the nonminimally coupled scalar field arise in the form of extra terms in the metric perturbations. Such a nontrivial scalar configuration may induce observable physical consequences, including modifications of test particle motion relative to the standard GR case. Substituting Eq. (21) into Eqs. (19) and (20), we obtain
\begin{equation}
	g'_{00}=-1+\frac{2m}{G_4(0)r'}(1+\frac{e^{-m_sr'}}{\zeta})-\frac{K(0)r'^2}{6G_4(0)}\Big(1-\frac{2}{\zeta}\Big)
\end{equation}
\begin{equation}
	g'_{ij}=\delta_{ij}\Big[1+\frac{2m}{G_4(0)r'}(1-\frac{e^{-m_sr'}}{\zeta})+\frac{K(0)r'^2}{12G_4(0)}\Big(1-\frac{4}{\zeta}\Big)\Big]
\end{equation}

\begin{equation}
	\phi'=\phi_0\Big(1+\frac{2m\chi e^{-m_sr'}}{\phi_0r'}+\frac{K(0)\chi r'^2}{3\phi_0}\Big)
\end{equation}
where 
\begin{equation}
	\frac{1}{\zeta}=\chi G_{4,\phi}(0)=\frac{G^2_{4,\phi}(0)}{G_4(0)[K_{,X}(0)-2G_{3,\phi}(0)+3G^2_{4,\phi}(0)/G_4(0)]}
\end{equation}
As seen from Eqs. (22-24), the Yukawa-type solution arises from the linearization of the Horndeski field equations. The terms proportional to \(\bar{x}_i^2\) appearing in the Cartesian representation of the weak-field solution do not correspond to a physical anisotropy of the spacetime. They arise from the coordinate form of the solution before imposing isotropic coordinates. After applying the isotropic coordinate transformation, these contributions are absorbed into the radial part of the metric and the spatial part becomes proportional to \(\delta_{ij}\), as shown in Eq.(20). In the following section, the isotropic metric will be further rewritten in Schwarzschild-like coordinates for the analysis of geodesic motion and Solar-System observables.
Moreover, the properties of the Yukawa-type corrections are controlled by the parameter $\zeta$. This parameter is completely determined by the background values of the Horndeski functions K, $G_3$ and  $G_4$, together with their derivatives with respect to $\phi$ and X, evaluated at the background values. In particular, it involvesthe background values  $G_4(0)$ and its derivative $G_{4,\phi}(0)$ , which describe the coupling between the scalar field and the geometry, and the kinetic structure through $K_{,X}(0)$ and the scalar–tensor mixing term $G_{3,\phi}(0)$. Thus, it reflects the scalar–geometry coupling and the braiding interaction, while remaining independent of the explicit values of K(0) and $G_3(0)$. 
 
 To analyze the weak-field behavior of the theory within the linearized Horndeski framework, we derive the effective PPN parameters. The parameter $\gamma$ takes the form
\begin{equation}
	\gamma=\frac{{1-\frac{e^{-m_sr}}{\zeta}}}{{1+\frac{e^{-m_sr}}{\zeta}}}
\end{equation}
The above expression should be regarded only as an effective PPN-like parameter defined within the linearized weak-field approximation. Since screening mechanisms are not incorporated in the present analysis, it should not be identified with the physical PPN parameter measured in Solar-System experiments, but instead characterizes linearized deviations from General Relativity. The effective parameter $\gamma$ depends on the scalar-field mass $m_s$ and the parameter $\zeta$, which is determined by the Horndeski functions $K$, $G_3$, and $G_4$, together with their derivatives with respect to $\phi$. In the limit of a sufficiently massive scalar field, the Yukawa correction becomes exponentially suppressed and the General Relativity value $\gamma=1$ is recovered. The standard PPN formalism is conventionally formulated for asymptotically flat spacetimes, whereas the present weak-field solution contains effective curvature corrections proportional to $K(0)r^2$. However, within the adopted local Solar-System weak-field regime, the quantity $|K(0)|r^2/G_4(0)$ remains perturbatively negligible. Consequently, the local geometry relevant for Solar-System observations remains very close to Minkowski spacetime, and the effective parameter $\gamma$ still provides a meaningful characterization of linearized deviations from GR at observational scales. 

In the BDV theory, the parameter $\zeta$ exhibits a similar dependence through the Brans-Dicke parameter $\omega$. In the appropriate limit, the Horndeski expression for $\zeta$ reduces to the corresponding BDV form \cite{Hatice2}. The resulting expression for the effective parameter $\gamma$ enables a qualitative comparison with the observational precision achieved by the Cassini experiment, which constrains deviations from $\gamma=1$ at the level $\gamma_{\mathrm{observed}}-1=(2.1\mp2.3)\times10^{-5}$ \cite{Bertotti}. Accordingly, the Cassini result is employed here only as an estimate of the observational sensitivity to linearized deviations from GR, rather than as a strict bound on the full nonlinear theory. In this context, different physical regimes may lead to $\gamma$ approaching unity, depending on the scalar field mass  $m_s$ and its coupling to matter \cite{Will3,Alsing-berti-will,Perivolaropoulos}. One possible regime is the limit $e^{-m_sr}$$\to 0$, which corresponds to a very massive scalar field, i.e., $m_s \to\infty$. Since the scalar field acquires a large mass in this regime, its influence on gravitational interactions is exponentially suppressed and $\gamma\sim 1$ regardless of Horndeski functions. On the other hand, if the scalar mass is not sufficiently large and the exponential factor $e^{-m_sr}$ remains of order unity, corresponding to a very light scalar field, (i.e.,  $m_s \to 0$), then  $\gamma(r) $ reduces to the constant value $\gamma=\frac{\zeta-1}{\zeta+1}$.

The quantity $G_4(0)$ represents the value of the function $G_4(\phi)$ evaluated at the background scalar-field value, and it plays a crucial role in determining the effective strength of gravity in the weak-field limit, since $G_4(\phi,X)$ couples directly to the Ricci scalar in the action. Given the two limiting cases discussed above, it becomes essential to properly fix $G_4(0)$ to ensure consistency with the Newtonian limit. This requirement is met  by analyzing the mass dependent term in  $g_{00}$ component of the metric. In the case of a very massive scalar field, i.e., where $m_s\gg 1$, the exponential decay of the Yukawa-type term ensures that its contribution becomes negligible at relevant scales, so we can take $G_4(0)=1$. In contrast, for a light scalar field, i.e. $m_s\ll 1$, a comparison with the Newtonian potential shows that $G_4(0)=\frac{\zeta+1}{\zeta}$. 
Taking the observational sensitivity of the Cassini experiment as a reference scale, linearized deviations associated with $\zeta \gtrsim 8\times10^4$ are not expected to produce observable effects at Solar-System scales within the present weak-field approximation. Accordingly, the Cassini sensitivity may be used to estimate the parameter region in which the linearized Horndeski corrections remain observationally indistinguishable from the General Relativity prediction. Here $\zeta$ denotes an effective model-dependent coupling parameter determined by specific combinations of the Horndeski functions and their derivatives evaluated at the background scalar-field value. Within the adopted weak-field approximation, the Solar-System tests analyzed in the present work may therefore be used to place approximate bounds on effective quantities such as the scalar-field mass \(m_s\) and the coupling parameter \(\zeta\) associated with small deviations from General Relativity. However, such estimates remain model-dependent and should be interpreted only within the perturbative approximation employed in the present analysis.
In the case of intermediate scalar masses, the previous method is insufficient to determine $G_4(0)$, but it is possible to introduce an effective gravitational coupling that incorporates the exponential function,
\begin{equation}
	G(r)=\frac{1}{G_4(0)}\Big(1+\frac{e^{-m_s r}}{\zeta} \Big).
\end{equation}
A detailed numerical treatment of this scenario is given in \cite{Perivolaropoulos}, where the PPN approach assumes asymptotic flatness and thus requires the background potential to vanish, in contrast to the approach adopted in our study.
\section{Perihelion Advance in the Linearized Regime}  
In this section, the effects of both a point mass and scalar field potential on the motion of test particle are analyzed, with particular emphasis on how the scalar potential modifies the weak-field spacetime geometry and affects the geodesic trajectories within the linearized framework. Given its impact on the geodesic structure in the linearized regime, we begin by computing the perihelion precession of test particles induced by the linearized scalar–tensor background of Horndeski theory with a potential. Subsequently, we extend our analysis to other classical Solar System tests, such as light deflection and gravitational redshift, within the same linearized framework, where nonlinear screening effects are neglected. For consistency with previous works and in order to analyze geodesic motion, it is useful to rewrite the point-mass solution in a compact isotropic form before transforming it to Schwarzschild-like coordinates. Using the scalar-field solution $\varphi'(r')$, the metric components $g'_{00}$ and $g'_{ij}$ can be equivalently expressed as
\begin{equation}
	ds^2 =
	-\left(
	1-\frac{2m}{G_4(0)r'}
	+\frac{K(0)}{6G_4(0)}r'^{\,2}
	-\sigma\varphi'
	\right)dt'^{\,2}
	+
	\left(
	1+\frac{2m}{G_4(0)r'}
	+\frac{K(0)}{12G_4(0)}r'^{\,2}
	-\sigma\varphi'
	\right)
	\left(dr'^{\,2}+r'^{\,2}d\Omega^2\right).
\end{equation}
The following radial coordinate transformation is introduced in order to bring the angular part of the metric into the standard Schwarzschild-like form:
\begin{equation}
	r'=r\left(
	1-\frac{K(0)}{24G_4(0)}r^2
	+\frac{\sigma\varphi}{2}
	\right).
\end{equation}
Since both the radial coordinate and its differential transform under this redefinition, and since the scalar perturbation depends on the radial coordinate, the radial metric component receives additional contributions involving derivatives of the scalar field. Introducing
\begin{equation}
	\alpha(r)=\frac{d\varphi}{dr},
\end{equation}
The derivative contribution appearing in the radial metric component is not merely a coordinate effect, but arises from the radial dependence of the scalar perturbation within the perturbative transformation. The metric takes the form
\begin{equation}
	ds^2=-\Big(1-\frac{2m}{G_4(0)r}+\frac{K(0)r^2}{6G_4(0)}-\sigma\varphi\Big)dt^2+\Big(1+\frac{2m}{G_4(0)r}-\frac{K(0)r^2}{6G_4(0)}+\sigma\alpha r\Big)dr^2+r^2 d\Omega^2 .
\end{equation}
where $\varphi$(r) is given by
\begin{equation}
	\varphi({r})=\frac{2m\chi}{r}e^{-m_sr}+\frac{K(0)\chi r^2}{3}.
\end{equation}
The appearance of the term $\sigma\alpha r$ in the radial metric component originates from the derivative contributions generated by the scalar perturbation under the radial coordinate transformation. Unlike the vacuum Schwarzschild solution of General Relativity, the scalar degree of freedom contributes independently to the time and radial sectors of the metric. As a result, the Schwarzschild relation $g_{tt}g_{rr}=-1$ is not generally recovered in the present scalar-tensor theory, even within the weak-field approximation. This behavior reflects the fact that the gravitational interaction is mediated not only by the metric tensor but also by the additional scalar degree of freedom. Nevertheless, since the Schwarzschild-like metric is obtained perturbatively from the isotropic weak-field solution through a coordinate transformation, it continues to satisfy the linearized field equations at the same perturbative order.

A general static, spherically symmetric metric in Schwarzschild-type coordinates is given by
\begin{equation}
	ds^2=-A(r)dt^2+B(r)dr^2+r^2(d\theta^2+sin^2{\theta}d\Phi^2).
\end{equation}
Since the spacetime is spherically symmetric, the motion is confined to a single plane. Choosing this plane to be the equatorial one ($\theta$ = $\pi$/2) simplifies the equations. The motion of test particles or photons in a static, spherically symmetric spacetime can be described by the following Lagrangian:
\begin{equation}
	2\mathcal{L}=-A(r)\dot{t}^2+B(r)\dot{r}^2+r^2\dot{\Phi}^2.
\end{equation}
where the dot denotes differentiation with respect to proper time $\tau$ for massive (timelike) particles or with respect to an affine parameter for massless particles. Due to the time-translation and rotational symmetries of the spacetime, the corresponding conserved quantities denoted by E for specific energy and L for angular momentum, are given by
\begin{equation}
	\dot{t}=-\frac{E}{A(r)}  \qquad  \dot{\Phi}=\frac{L}{r^2}.
\end{equation}
Thus, the radial component of the geodesic equation takes the form
\begin{equation}
	\dot{r}^2=\frac{1}{B(r)}\Big(\kappa+\frac{E^2}{A(r)}-\frac{L^2}{r^2}\Big)
\end{equation}
Here, $\kappa$ is a normalization constant that takes the value -1 for timelike geodesics and 0 for null geodesics.
To derive the orbit equation, we divide the radial equation by the square of the angular velocity $\dot{\Phi}^2$, which leads to
\begin{equation}
	\Big(\frac{dr}{d \Phi}^2\Big)=r^4\Big(\frac{\kappa}{L^2 B(r)}+\frac{E^2}{L^2 A(r)  B(r)}-\frac{1}{B(r)r^2}\Big)
\end{equation}
To simplify the orbital differential equation, we introduce the inverse radial coordinate given by
\begin{equation}
	u=\frac{1}{r}
\end{equation}
Using the relation
\begin{equation}
	\frac{dr}{d\Phi}
	=
	-\frac{1}{u^2}\frac{du}{d\Phi},
\end{equation}
the radial geodesic equation can be rewritten entirely in terms of the variable \(u\). In particular, the Yukawa-type factor appearing in the scalar contribution transforms as $e^{-m_s r}=e^{-m_s/u}$. By differentiating and using Eq. (31), one obtains the following differential equation.
\begin{equation}
	\begin{aligned}
		\Big( \frac{du}{d\Phi} \Big)^2 + u^2
		&= \Big( \frac{E^2 - 1}{L^2} \Big)
		+ \frac{2mu^3}{G_4(0)}
		+ \frac{2mu}{G_4(0)L^2}
		+ 2m\chi\sigma u e^{-m_s/u}
		\Big( \frac{2E^2 - 1}{L^2} \Big) \\[6pt]
		&\quad
		- \frac{K(0)}{6G_4(0)}
		\Big[ 1 - 4\chi G_{4,\phi}(0) \Big]
		- \frac{K(0)}{6G_4(0)}u^{-2}
		\Big[1 + 2\chi G_{4,\phi}(0)(E^2-2) \Big]
	\end{aligned}
\end{equation}
By applying differentiation with respect to $\Phi$, the equation transforms into a modified form of the Binet equation
\begin{equation}
	\frac{d^2 u}{d\Phi^2}+u=\frac{m}{L^2G_4(0)}+\frac{3m u^2}{G_4(0)}+\frac{K(0) }{6G_4(0) L^2u^3}\Big(1+2\chi G_{4,\phi}(0)(E^2-2)\Big)+\frac{m\chi\sigma }{L^2}e^{-m_s/u}\Big[2E^2-1-3L^2u^2+O(m_s,m_s^2)\Big]
\end{equation}
As discussed previously, we consider two situations. Firstly we investigate for the very massive scalar field $m_s\gg1$ and $G_4(0)=1$
\begin{equation}
\frac{d^2 u}{d\Phi^2}+u=\frac{m}{L^2G_4(0)}+\frac{3m u^2}{G_4(0)}+\frac{K(0) E^2}{6G_4(0) L^2u^3}\Big(1+2\chi G_{4,\phi}(0)(E^2-2))\Big)
\end{equation}
 In this paper, the Adkins–McDonnell one-dimensional integral formulation, as simplified by Arakida, is adopted to compute perihelion precession under arbitrary central-force perturbations \cite{arakida-perihelion}. Specifically, the impact of a scalar-field potential on the resulting precession is analyzed. The integration formula is given by
\begin{equation}
	\Delta=\frac{-2l}{me^2}\int_{-1}^{1}\frac{zdz}{\sqrt{1-z^2}}\frac{d\tilde{V}(z)}{dz}
\end{equation}
where $ r=l/(1+ez)$ and $l=L^2/GM=a(1-e^2)$, a denotes the semi-major axis of the orbit, l represents the semi-latus rectum, and e characterizes the orbital eccentricity. More detailed discussions on the evaluation of this type of integral can be found in the related literature \cite{arakida-perihelion,adkins,Ovcherenko,chashchina}. The resulting orbital equation can be written as
\begin{equation}
	u=\frac{1}{r}=\frac{m}{L^2}[1+ecos(1-\epsilon)\Phi]
\end{equation}
where $\Delta=2\pi\epsilon$, $\epsilon$ represents the perturbative correction to the angular dependence of the orbit. As shown in Eq.(44), the expression contains the deviations from the Newtonian potential in central motion. These deviations, denoted by $\tilde{V}(r)$, can be obtained from the complete potential of the corresponding gravity theory for a point mass, and can be written in the form,
\begin{equation}
	U(r)=\frac{-m}{r}+\frac{L^2}{2r^2}+\tilde{V(r)}
\end{equation}
We can find the $\tilde{V}(r)$ term using the radial geodesic equation, which yields
\begin{equation}
\frac{\dot{r}^2}{2}+U(r)=\frac{E^2}{2}
\end{equation}
This relation takes the form of an energy conservation law, where $\frac{\dot{r}^2}{2}$ corresponds to the radial kinetic term, U(r) denotes the effective potential. Accordingly, the effective potential takes the form
\begin{equation}
	U(r)=\frac{1}{2}\Big(1+\frac{L^2}{r^2}\Big)\Big(1-\frac{2m}{G_4(0)r}+\frac{K(0)r^2}{6G_4(0)}-\sigma\alpha r\Big)+\frac{E^2}{2}\sigma(\alpha r-\varphi)
\end{equation}
Using the values of $\alpha$ and $\varphi$ from Eqs.(30) and (32) and  performing the necessary algebra, we obtain the following expression
\begin{equation}
	U(r)=\frac{1}{2}\Big(1+\frac{L^2}{r^2}\Big)\Big[1-\frac{2m}{G_4(0)r}\Big(1-\chi G_{4,\phi}(0) e^{-m_s r}\Big)+\frac{K(0)r^2}{6G_4(0)}\Big(1- 4\chi G_{4,\phi}(0) \Big)\Big]-\frac{2 m \chi \sigma E^2  e^{-m_s r}}{r}+ \frac{K(0) E^2\chi \sigma r^2 }{6}
\end{equation}
By comparing Eq.(45) with the explicit expression for the effective potential, the correction term $\tilde V(r)$ can be written as
\begin{equation}
	\tilde V(r)=
	-\frac{mL^2}{G_4(0)r^3}
	\left(1-\chi G_{4,\phi}(0)e^{-m_s r}\right)
	+\frac{K(0)r^2}{12G_4(0)}
	\left(1-4\chi G_{4,\phi}(0)\right)
	+\frac{K(0)L^2}{12G_4(0)}
	\left(1-4\chi G_{4,\phi}(0)\right)
	-\frac{2m\chi\sigma E^2e^{-m_s r}}{r}
	+\frac{K(0)\chi\sigma E^2r^2}{6}.
\end{equation}
The term proportional to $K(0)L^2$ is independent of $r$ and therefore does not contribute to the perihelion advance. In addition, the Yukawa-type $1/r$ term is absorbed into the effective Newtonian contribution and is not included in the perturbing potential used for the perihelion calculation. Moreover, for non-relativistic bound orbits we take $E^2\simeq 1$. Thus,
\begin{equation}
	\tilde V(r)\rightarrow
	-\frac{mL^2}{G_4(0)r^3}
	\left(1-\chi G_{4,\phi}(0)e^{-m_s r}\right)
	+\frac{K(0)r^2}{12G_4(0)}
	\left(1-4\chi G_{4,\phi}(0)\right)
	+\frac{K(0)\chi\sigma r^2}{6}.
\end{equation}
Because the potential contains exponential contributions of the form $e^{-m_s r}$, the corresponding integrals cannot, in general, be solved analytically. It is therefore convenient to study the  very heavy and light scalar field regimes separately, whereby analytic results for the perihelion advance can be derived. In the heavy scalar field case, the Yukawa contribution is exponentially suppressed, i.e. \(e^{-m_s r}\rightarrow 0\). In this limit, we set $G_4(0)=1$ and use $\frac{1}{\zeta}=\chi G_{4,\phi}(0)$. Accordingly, the effective correction potential reduces to
\begin{equation}
	\tilde{V(r)}_{m_{s}\rightarrow\infty}=-\frac{mL^2}{r^3}+\frac{K(0)r^2}{12}\Big(\frac{\zeta-2}{\zeta}\Big)
\end{equation}
Because the result of integral (43) is obtained for power-law potentials \cite{adkins,arakida-perihelion}, the only difference compared with their result arises from the $\zeta$ dependent coefficient appearing in  $\tilde V(r)$. Therefore, the resulting expression can be written as
\begin{equation}
	\Delta=\Delta_E-\Big(\frac{\zeta-2}{\zeta}\Big)\Delta_{K_0}
\end{equation}
In order to make this relation more explicit in terms of the orbital parameters, we introduce the following definitions
\begin{equation}
	\Delta_E=\frac{6\pi m}{a(1-e^2)}..
\end{equation}
\begin{equation}
	\Delta_{K_0}=-\frac{\pi K_{0}}{2m}a^3\sqrt{1-e^2}
\end{equation}
According to the result (50), in the heavy-scalar limit, Yukawa-type terms become exponentially suppressed and the perihelion advance reduces to two components. Here, $\Delta_E$ denotes the standard general relativistic correction to orbital motion, arising from the central mass and orbital geometry, while the contribution
$\Delta_{K_0}$, representing the background curvature or cosmological term $K_0$, is rescaled within the scalar–tensor framework by the factor $C(\zeta)=\Big(\frac{\zeta-2}{\zeta}\Big)$. Under the assumption $K(0)=-2\Lambda G_4(0)$, this term yields the contribution of the cosmological constant, expressed as $\Delta_\Lambda=\frac{\pi \Lambda}{m}a^3\sqrt{1-e^2}$. The result has a similar structure to the GR$\Lambda$ theory, with the distinction arising from the $\zeta$ dependent numerical coefficients multiplying the second term. Since the scalar field is exponentially suppressed at sufficiently large distances in the linearized massive Horndeski model, its contribution becomes negligible, and the expression approaches the General Relativity result. When we examine the contribution coming from the minimum of the potential, we find that when the minimum of the potential is zero, the perihelion precession expression approaches the General Relativity result, showing no dependence on the parameter $\zeta$. However, if the minimum of the potential $K_0$ is not zero, the resulting perihelion shift contains an additional term involving a factor of $\zeta$, unless $\zeta = 2$. Since Solar-System measurements become less sensitive to the parameter $\zeta$ in the very heavy scalar-field limit due to Yukawa suppression, the second term may lead to deviations from the GR$\Lambda$ expression for certain parameter choices. Therefore, for the case when the mass of the scalar field is very heavy, the effect of the minimum of the potential may differ from the corresponding GR result, even when the same observed value of the cosmological constant is used. Although the effect of the potential is significantly weaker than the classical Einstein correction, it still provides a valuable theoretical perspective on how it can manifest itself in local gravitational systems. If future observations were to determine $K_0$, constraints on $\zeta$ could be derived within the heavy scalar field regime of Horndeski theory considered here.
For a very light scalar field, the Yukawa factor can be expanded as
$e^{-m_s r}\simeq 1-m_s r$. Neglecting terms proportional to $mm_s$, and using
$G_4(0)=\frac{\zeta+1}{\zeta}$ together with
$\chi G_{4,\phi}(0)=\frac{1}{\zeta}$, we obtain
\[
-\frac{mL^2}{G_4(0)r^3}
\left(1-\chi G_{4,\phi}(0)e^{-m_s r}\right)
=
-\frac{mL^2}{r^3}
\frac{\zeta}{\zeta+1}
\left(1-\frac{1}{\zeta}\right)
=
-\left(\frac{\zeta-1}{\zeta+1}\right)
\frac{mL^2}{r^3}.
\]
Therefore,
\begin{equation}
	\tilde V(r)=
	-\left(\frac{\zeta-1}{\zeta+1}\right)\frac{mL^2}{r^3}
	+\left(\frac{\zeta-2}{\zeta+1}\right)\frac{K(0)r^2}{12}.
\end{equation}
This potential leads to an additional contribution to the perihelion shift, obtained in the form
\begin{equation}
	\Delta=\Big(\frac{\zeta-1}{\zeta+1 }\Big)\Delta_E-\Big(\frac{\zeta-2}{\zeta+1}\Big)\Delta_{K_0}
\end{equation}
In the limit  $\zeta\to \infty$, the coefficients tend to unity and the expression reduces to $\Delta_E- \Delta_{K_0}$. Since $\Delta_{K_0}$ represents the contribution from the potential minimum, this corresponds not to pure General Relativity but rather to General Relativity with a cosmological constant (GR$\Lambda$). However, when the potential vanishes, the large $\zeta$ regime leads to expressions that reduce to  the General Relativity result, $\Delta \to \Delta_E$. At finite values of  $\zeta$, especially when  $\zeta$ is small, the deviations become substantial. The Einstein term vanishes at $\zeta=1$, while the potential contribution disappears at $\zeta=2$. Thus, only in the large-\(\zeta\) regime do the derived expressions approach the GR\(\Lambda\)-like limit when \(K(0)\neq0\), while they reduce to the standard GR result when the potential contribution vanishes. For smaller values of \(\zeta\), the very light scalar field can generate appreciable deviations from these limits. Using the Cassini sensitivity as a reference, the regime $\zeta\gtrsim 8\times10^4$ corresponds to linearized scalar-field corrections that are expected to remain observationally negligible at Solar-System scales. However, when the potential vanishes, the large $\zeta$ regime leads to expressions that reduce to the General Relativity result, $\Delta \to \Delta_E$. At finite values of $\zeta$, especially when $\zeta$ is small, the deviations become substantial. The Einstein term vanishes at $\zeta=1$, while the potential contribution disappears at $\zeta=2$.

In conclusion, as seen from Eqs. (52) and (56), the contributions to the advance of perihelion due to the mass of the source are similar in form to the corresponding GR results. It should be noted that the multiplicative factors accompanying the mass term play the role of effective PPN-like parameters within the adopted linearized framework. In the light scalar-field regime, sufficiently large values of $\zeta$ suppress the scalar-mediated corrections, so that the derived expressions approach the General Relativity limit at Solar-System scales. For a very massive scalar field, however, this contribution becomes independent of $\zeta$ and approaches the corresponding Genral Relativity value. Therefore, the contribution arising from the mass of the source has a similar form to that of the corresponding GR expression for all values of $\zeta$. When we take into account the effects due to the  minimum of the potential, the resulting terms are structurally similar to the corresponding GR$\Lambda$ expressions. The differences arise from the multiplicative factors involving the term $\zeta$. The coefficient in front of the potential term, $C(\zeta)=\Big(\frac{2-\zeta}{\zeta}\Big)$ plays a central role in governing deviations from the General Relativity limit. The system exhibits distinct behaviors for different values of $\zeta$. Depending on the sign and magnitude of $\zeta$, the coefficient $C(\zeta)$ can change both sign and amplitude, leading to either an enhancement or a suppression of the potential term contribution to the perihelion precession. Consequently, future high-precision measurements of the perihelion precession may help clarify the role of the parameter $\zeta$ in the weak-field regime analyzed in this work.

\section{Deflection of Light Rays in the Linearized Regime}    
In this section, we analyze deflection of light rays within the linearized massive Horndeski theory with a potential, and examine how both the potential minimum and the scalar field mass affect this process. In this work, as in previous studies, the Schwarzschild-like coordinates  given in Eq.(31) are adopted. The orbit equation becomes,         
\begin{equation}
	\frac{d^2 u}{d\Phi^2}+u=\frac{3m u^2}{G_4(0)}+\frac{K(0) \chi\sigma E^2}{3L^2u^3}+\frac{m\chi\sigma }{L^2}e^{-m_s/u}\Big[2E^2-3L^2u^2+O(m_s,m_s^2)\Big]
\end{equation}
The above expression for the orbit contains Yukawa-type terms, whose behavior depends on the scalar field mass. As in the previous calculations, the cases of a very heavy scalar field ($m_s \to \infty$) and and a very light scalar field ($m_s \to 0$) are analyzed seperately to examine their physical implications. The expression is first evaluated in the limit of a very heavy scalar field where the exponential contributions become negligible. Setting $G_4(0)=1$ under this approximation, Eq. (57) reduces to
\begin{equation}
	\frac{d^2 u}{d\Phi^2}+u=3m u^2+\frac{K(0)\chi\sigma}{3b^2u^3}
\end{equation}
Here, $b=L/E$ denotes the impact parameter associated with the unperturbed photon trajectory. The first term on the right-hand side corresponds to the classical gravitational deflection term arising from the Schwarzschild geometry in General Relativity (GR), while the second term represents the contribution associated with the minimum of the scalar field potential. Hence, the minimum of the scalar field potential enters the light-deflection formula as an additional correction. We analyze  the linear-order solution of the orbit equation and apply a perturbative method to investigate the effect of the point mass and the scalar field potential on a light ray coming from the distant region of spacetime, with the following ansatz
\begin{equation}
	u(\Phi)=u_0(\Phi)+ m u_1(\Phi)+K(0)u_2(\Phi)
\end{equation}
Substituting equation (59) into equation (58), we obtain three equations corresponding to the zeroth order and the terms linear in m and K(0). At zeroth order, the solution corresponds to a photon following a straight-line path in the background spacetime, which is given by
\begin{equation}
	u_0(\Phi)=\frac{sin\Phi}{R}
\end{equation}
where $R$ is an integration constant arising from the perturbative solution of the orbit equation. By inserting this expression into the remaining equations, the following solution is obtained
\begin{equation}
	u(\Phi)=\frac{1}{r}=\frac{sin\Phi}{R}+\frac{3m}{2R^2}\left(1+\frac{1}{3}cos2\Phi)\right)+\frac{K(0)\chi\sigma R^3cos^2\Phi}{6b^2sin\Phi}
\end{equation}
 This equation provides an approximate solution for the path of light in a gravitational field. The first term represents the straight line motion of light in flat spacetime where no gravitational effects are present. The second term represents the general relativistic correction to the light’s trajectory, arising from the curvature of spacetime produced by a massive object. The third term corresponds to an additional geometric correction in the light trajectory induced by the minimum of the scalar-field potential. From the relation $K(0)=-2\Lambda G_4(0)$, it follows that negative values of $K(0)$ correspond to a positive effective cosmological constant, leading to de Sitter-like curvature corrections within the weak-field regime, whereas positive values of $K(0)$ correspond to an effective negative cosmological constant associated with anti–de Sitter–like curvature contributions. Moreover, due to its dependence on $\Phi$, this term vanishes at the point of closest approach and becomes dominant in the asymptotic regions. This indicates that the deflection of light receives contributions not only by local gravitational effects but also from terms associated with the large-scale curvature of spacetime. We find the relation between the integration constant R and closest approach distance by imposing $\Phi=\pi/2$ in Eq.(61)
\begin{equation}
	\frac{1}{r_0}=\frac{1}{R}+\frac{m}{R^2}
\end{equation}
where $r_0$ denotes the distance of closest approach determined from the condition $dr/d\Phi=0$. Therefore, within the linearized weak-field approximation adopted throughout this work, the difference between $R$ and $r_0$ appears only at higher order in the perturbation parameters. Imposing the closest-approach condition $dr/d\Phi=0$ further yields
\begin{equation}
	\frac{1}{b^2}-\frac{K(0)}{6G_4(0)}(1-2\chi G_{4,\phi})=\frac{1}{r_0^2}-\frac{2m}{r_0^3}
\end{equation}
or equivalently
\begin{equation}
	\frac{1}{b^2}-\frac{K(0)}{6}\Big(\frac{\zeta-2}{\zeta}\Big)=\frac{1}{r_0^2}-\frac{2m}{r_0^3}
\end{equation}
Equations (62) and (63) show that the differences among (R), $r_0$, and (b) contribute only beyond linear order in the weak-field expansion. Therefore, within the accuracy of the present approximation, these quantities may consistently be treated as approximately equivalent. Consequently, the orbit equation and the corresponding bending-angle expression can be written in terms of a single parameter, chosen here as (R), which leads to the simplified form of Eq. (61).
\begin{equation}
	u(\Phi)=\frac{sin\Phi}{R}+\frac{3m}{2R^2}\left(1+\frac{1}{3}cos2\Phi\right)+\frac{K(0)\chi\sigma Rcos^2\Phi}{6sin\Phi}
\end{equation}
As demonstrated in the pioneering study \cite{Rindler-ishak}, light deflection receives contributions not only from the local geometry but also from the global geometry of spacetime, including the cosmological constant. This result arises because the spacetime with a nonzero cosmological constant is not asymptotically flat, and thus the classical Schwarzschild definitions of deflection and impact parameter must be reformulated accordingly. Subsequent studies \cite{Ishak-rindler,Rindler-ishak,Sereno,Schucker,Khriplovich,Park,light3,Islam,arakida1,Arakida2,Bhadra,Faraoni,Biressa,Lebedev-lake,Lebedev-lake1,Kim,Simpson,Patella} examined this problem in detail and revealed that the dependence of the bending angle on cosmological constant is not a universal feature but rather depends on how the angle is defined and measured, as well as on how physical quantities such as the impact parameter are specified. In the present study, the spacetime is not asymptotically flat due to the presence of a nonvanishing minimum potential. As a result, the classical method cannot be applicable, since the curvature of space-time persists at large distances. It should also be emphasized that the effective curvature contributions proportional to $K(0)r^2$ modify the asymptotic structure even within the linearized weak-field regime. Consequently, the light-deflection analysis must be formulated through local angular measurements that consistently incorporate the non-flat geometry. Therefore the Rindler-Ishak approach presented in \cite{Rindler-ishak,Ishak-rindler} is adopted and  generalized  to investigate the linearized Horndeski theory with a potential. In this analysis, both the source and observer are assumed to be static. The bending angle is calculated using the invariant formula for the cosine of the angle between the coordinate directions d and $\delta$, as illustrated Fig.1, which is given by
\begin{equation}
	 \cos\psi=g_{ij} d^i \delta^j/[\sqrt{g_{ij}d^i d^j}\sqrt{g_{ij}\delta^i \delta^j}]
\end{equation}
Here $d=(dr,d\Phi)=(\dot{r},1)d\Phi$ with $\beta=dr/d\Phi$ denotes the direction of the photon’s orbit and $\delta=(\delta r, 0)$ represents the direction of the coordinate line $\Phi=\mbox{constant}$. 
\begin{figure}[h]
	\begin{centering}
		\includegraphics[draft=false, width=7cm]{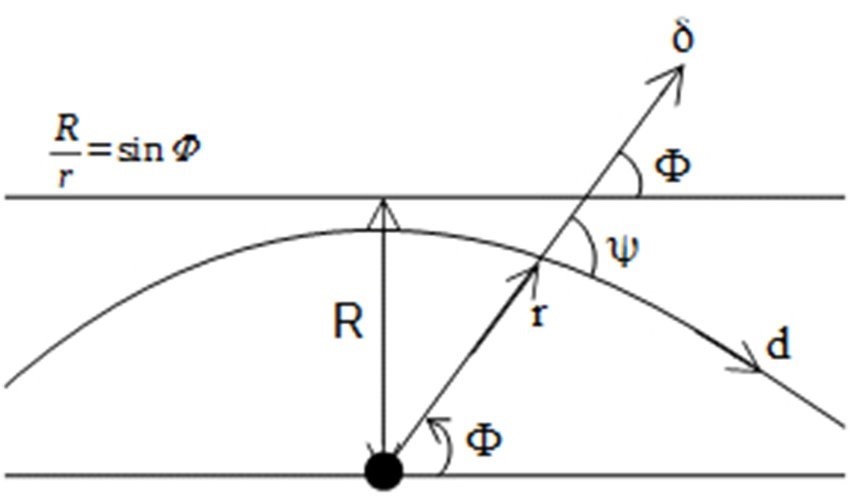}
		\caption{The plane graph  corresponding to the orbit  equation given in equation (3). 
			The one-sided deflection angle is given by $\alpha=\psi-\Phi$ (The figure is adapted from \cite{Rindler-ishak}).}
		\label{f1}
	\end{centering}
\end{figure}
The metric component $g_{ij}$ corresponds to the two-dimensional submanifold of  the metric (31), defined by t=constant, $\theta=\pi/2$. Then, the parameter  $\beta$ is obtained as follows
\begin{equation}
	\beta=\frac{dr}{d\phi}=-\frac{r^2}{R}cos\Phi+\frac{mr^2}{R^2}sin2\Phi+\frac{K(0)\chi\sigma Rcos\Phi}{6}\Big(1+\frac{1}{sin^2\Phi}\Big)
\end{equation}
From the above relations, it follows that $\cos\psi= |\beta|/\sqrt{\beta^2+r^2/B}$. Using $\tan\psi=\sqrt{\sec^2\psi-1}$  we obtain
\begin{equation}
	tan\psi=\frac{r}{\vert\beta\vert \sqrt{B}}
\end{equation}
Since the solutions exhibit a singularity at  $\Phi=0$, the deflection angle is measured at $\Phi=\Phi_0\ll 1 $, where the gravitational deflection due to mass has already been realized. In this regime, the small-angle approximations $\sin\Phi_0\approx \Phi_0$ and  $\cos\Phi_0\approx 1$  are therefore employed. In accordance with the weak-field and small-angle approximation, $\Phi_0$ is taken to be order $\Phi_0\approx O(m/R)$. Thus Eq.(61) takes the following form
\begin{equation}
	u=\frac{1}{r}=\frac{\Phi_0}{R}+\frac{2m}{R^2}++\frac{K(0)\chi\sigma R}{6\Phi_0}
\end{equation}
Accordingly, Eq.(67) can be written as follows
\begin{equation}
	\abs{\beta}=\frac{r^2}{R}\Big[1-\frac{2m}{R}\Phi_0-\frac{K(0)\chi\sigma R^2}{6}\Big(1+\frac{1}{\Phi_0^2}\Big)\Big]
\end{equation}
 The metric function B(r) evaluated at the corresponding r value and the parameter $\beta$ given in Eq.(70) are substituted into eq.(68), which gives the following result up to linear order in m, K(0) and $\Phi_0$.
\begin{equation}
\psi \approx \Phi_0+\frac{2m}{R}+\frac{K(0)\chi\sigma R^2}{6\Phi_0}+\frac{K(0)R^3(1-4\chi G_{4,\phi}(0))}{12\Big[\Phi_0 R+2m+\frac{K(0)\chi\sigma R^3}{6\Phi_0}\Big]}+\frac{K(0)\chi \sigma R^3}{6\Phi_0^2}\Big[\frac{\Phi_0 R+2m+\frac{K(0)\chi\sigma R^3}{6\Phi_0}}{R^2}\Big]
\end{equation}
In the linearized order, the one-sided bending angle $\alpha=\psi-\Phi$ is obtained using Eq.(25) as follows
\begin{equation}
	\alpha \approx \alpha_E+\frac{K(0) R^2}{6\zeta \Phi_0}+\frac{K(0) R^2}{6\zeta}\Big[\frac{(\zeta-4)r}{2R}+\frac{R}{\Phi_0^2r}\Big]
\end{equation}
In Eq.(72), the first term corresponds to the standard deflection angle predicted by classical General Relativity, representing the part of the deflection angle that is unaffected by the scalar-field modification in the linearized framework. The additional terms indicate the modifications induced by the scalar field potential on this classical result. The second term provides a constant correction to the deflection angle, determined by the characteristic scale of the system R, the coupling coefficient $\zeta$ and the minimum potential K(0). Being independent of the distance, this term represents a fixed contribution arising from the intrinsic parameters of the model.  The third term introduces a contribution that varies linearly with distance. When $\zeta=4$, the second component completely vanishes, and the additional deflection angle is described only by the constant and inversely proportional $1/r$ term. In contrast, when $\zeta\neq4$, the presence of the scalar field causes the deflection angle to increase or decrease linearly with distance, remains effective within a finite observational range. On the other hand, the final term  exhibits an inverse distance dependence, which enhances the scalar-induced contribution at short distances within the linearized framework and is suppressed at larger distances, where the result approaches the General Relativity limit.
We now consider the case of a light scalar field, for which the exponential terms in Eq.(57) can be expanded, leading Eq.(57) being expressed as
\begin{equation}
	\frac{d^2 u}{d\Phi^2}+u=3m u^2\Big(\frac{1}{G_4(0)}-\chi\sigma\Big)+\frac{K(0)\chi\sigma}{3b^2u^3}+\frac{2m\chi\sigma}{b^2}
\end{equation}
The solution to the orbit equation takes the form
\begin{equation}
u(\Phi)=\frac{1}{r}=\frac{sin\Phi}{R}+\frac{3m}{2R^2}\left(\frac{4\chi\sigma }{3}+\Big(\frac{1}{G_4(0)}-\chi\sigma\Big)\Big(1+\frac{cos 2\Phi}{3}\Big)\right)+\frac{K(0)\chi\sigma R^3\ cos^2\Phi}{6b^2sin\Phi}
\end{equation}
By rearranging this expression in accordance with Equation (25), we obtain the following result
\begin{equation}
	u(\Phi)=\frac{1}{r}=\frac{sin\Phi}{R}+\frac{3m}{2R^2(\zeta+1)}\left(\frac{4 }{3}+(\zeta-1)\Big(1+\frac{cos 2\Phi}{3}\Big)\right)+\frac{K(0) R^3\ cos^2\Phi}{6b^2sin\Phi(\zeta+1)}
\end{equation}
From this result, we see that K(0) enters the orbital equation and therefore contributes to the light deflection. The angle $\psi$, which represents the angle between the direction of the light ray at a given point and the radial direction, can be obtained as follows
\begin{equation}
	\psi \approx \Phi_0+\frac{2m}{R}\Big(\frac{ \zeta}{\zeta +1}\Big)+\frac{K(0)\chi\sigma R^2}{6\Phi_0}+\frac{K(0)R}{12}\Big(\frac{\zeta-4}{\zeta+1}\Big)\frac{R^2 }{\Big[\Phi_0 R++2m(\frac{\zeta}{\zeta+1})+\frac{K(0)\chi\sigma R^3}{6\Phi_0}\Big]}+\frac{K(0)\chi \sigma R^3}{6\Phi_0^2}\Big[\frac{\Phi_0 R+2m(\frac{\chi}{\chi+1})+\frac{K(0)\chi\sigma R^3}{6\Phi_0}}{R^2}\Big]
\end{equation}
To first order in m,$\Phi_0$ and K(0), the half bending angle can be written as
\begin{equation}
	\alpha \approx \frac{2m}{R}\Big(\frac{\zeta}{\zeta+1}\Big)+\frac{K(0) R^2}{6(\zeta+1)\Phi_0}+\frac{K(0) R^2}{6(\zeta+1)}\Big[\frac{(\zeta-4)r}{2R}+\frac{R}{\Phi_0^2 r}\Big]
\end{equation}	
The total deflection angle can be decomposed into the GR contribution and three additional terms arising from the scalar field. The Cassini result for the effective PPN parameter $\gamma$ indicates that the leading General Relativity contribution remains compatible with current observational sensitivity at Solar-System scales within the adopted linearized framework. The second term, originating from the scalar potential minimum $K(0)$ together with the scalar--matter couplings $(\chi,\sigma)$, provides an additional geometric contribution to the solution. Owing to its quadratic dependence on the characteristic scale $R$ and its suppression by the factor $1/(\zeta+1)$, this contribution is expected to remain small within the weak-field regime considered here, unless the effective curvature scale associated with $K(0)$ becomes sufficiently large.
 The last term represents a geometric correction to the light deflection angle in the light scalar field and reflects the contribution arising from the interaction between the scalar field and spacetime curvature. Moreover, it shows that, within the linearized framework, the background dynamics of the scalar field can contribute to modifications of light trajectories, even in regions without a local mass source. This term consists of two distinct physical contributions. The component proportional to r represents a contribution that increases with distance, becoming significant on cosmological scales while remaining negligible in local regions. In contrast, the term inversely proportional to r is associated with strong field variations near the source, reflecting the influence of local scalar field gradients. 
\section{Gravitational redshift in the Linearized Regime}
In this section, the gravitational redshift is calculated in the linearized regime in order to examine how the scalar field potential affects the frequency shift of photons propagating in the given spacetime geometry. As can be seen from the metrics (31) and (33), the timelike Killing vector gives rise to the standard gravitational redshift relation, which can be written as follows for the locally measured frequencies
\begin{equation}
	\frac{\nu_0}{\nu}=\sqrt{\frac{A(r)}{A(r_0)}}
\end{equation}
Here r and $r_0$ denote the radial positions of the observer and the emitter, respectively. Using the expression of A(r) given in (31), and keeping in mind that we are working at the linear approximation, equation (78) takes the form
\begin{equation}
\frac{\nu_0}{\nu}=1-\frac{m}{G_4(0)r}(1+\chi G_{4,\phi}(0)e^{-m_sr})+\frac{m}{G_4(0)r_0}(1+\chi G_{4,\phi}(0)e^{-m_sr_0})+\frac{K(0)}{12G_4(0)}(1-2\chi G_{4,\phi}(0))(r^2-r_0^2)
\end{equation}
In the given expression, the terms independent of the exponential term correspond to those appearing in the general relativistic limit of the model. They arise from the difference in the effective gravitational potential generated by the central mass, evaluated between the emission point r and the observation point $r_0$. In the GR limit, corresponding to a vanishing scalar field, the relation reduces to the classical form previously derived \cite{Kagramanova}. The multiplicative factor $\chi G_{4,\phi}(0)e^{-m_sr}$ introduces a Yukawa-type correction, indicating the presence of a massive scalar field. Within the linearized approximation, this term modifies the strength of the effective gravitational potential depending on the radial coordinate and the scalar-field mass. As a result, it  may either increase or decrease the gravitational redshift. The last term reflects a large-scale curvature generated by the scalar potential and acts as a $\Lambda$-like term in the linearized limit. Hence, the gravitational redshift is affected not only by the local gravitational potential difference but also by the effective background curvature produced by the scalar-field potential in the linearized regime. While the term proportional to the source mass involves a Yukawa-type exponential correction suppressed by the scalar-field mass $m_s$, the cosmological background term does not. Let us consider the two limiting cases of a heavy scalar field and a light scalar field to better understand their respective impacts on the redshift.
In the heavy scalar limit, $m_s \to \infty$, and setting $G_4(0)=1$, we obtain
\begin{equation}
	\frac{\nu_0}{\nu}=1-\frac{m}{r}+\frac{m}{r_0}+\frac{K(0)}{12}(1-2\chi G_{4,\phi}(0))(r^2-r_0^2)
\end{equation}
The Yukawa-type corrections in Eq.(79), associated with the local potential, decay exponentially and effectively vanish in this limit. Hence, for a heavy scalar field, the gravitational redshift due to mass reduce to the standard GR prediction, whereas the cosmological background contribution remains present, as it is not subject to Yukawa suppression. Therefore, the scalar field continues to rescale the background term even in the heavy limit. The multiplicative factor $(1-2\chi G_{4,\phi}(0))$ in the last term determines the strength of the coupling between the scalar field and the metric. This coefficient rescales the amplitude of the K(0) and therefore modifies the way the scalar field potential contributes to the redshift, determining whether it is strengthened, weakened, or completely cancelled in the special case  $\chi G_{4,\phi}(0)=1/2$. Although the last term has an explicit radial dependence, its variation across Solar-System distances is negligible, so its contribution to the redshift can be treated as effectively constant. While this contribution does not lead to measurable gravitational redshift effects at Solar-System scales, it represents a background geometric term whose magnitude increases with distance and becomes relevant only when approaching cosmological scales.
 We now turn to the opposite case, namely the very light scalar field with
 $m_s \to 0$ and discuss its implications. In this case  one may expand Yukawa factor as $e^{-m_s r} \sim 1-m_s r$. Neglecting terms proportional to $ m m_s$  and using the value of  $G_4(0)$ for a very light scalar field case, we obtain
\begin{equation}
	\frac{\nu_0}{\nu}=1-\frac{m}{r}+\frac{m}{r_0}+\frac{K(0)}{12G_4(0)}(1-2\chi G_{4,\phi}(0))(r^2-r_0^2)
\end{equation}
The first three terms arise directly from the Schwarzschild metric and describe the mass–dependent gravitational redshift, reproducing GR result, as expected in the heavy scalar field limit. Similar consistency also holds in BD$\Lambda$ or BDV theories \cite{Hatice2}. With $\frac{1}{\zeta}=\chi G_{4,\phi}(0)$, the coefficient $1-2\chi G_{4,\phi}(0)$ becomes $1-\frac{2}{\zeta}$ and therefore tends to unity for the large $\zeta$ values required in the light scalar-field regime. The effect of the potential, whose prefactor depends strongly on theory-specific parameters, exhibits a quadratic dependence on the radial separation, in contrast to the inverse radial behavior of the General Relativistic term. This result highlights the importance of extending phenomenological studies of scalar–tensor theories beyond local tests. At local Solar System scales, within the weak-field phenomenological framework considered here, this contribution is expected to be either strongly suppressed or effectively negligible. Consequently, although the potential contribution is formally present in the field equations, it becomes effectively indistinguishable from the GR term in the weak-field regime, as in BDV and $BD\Lambda$ theories where the prefactor also approaches unity \cite{Hatice2}.

\section{Conclusions}

In this paper, weak-field solutions of Horndeski theory in the Jordan frame are derived within the linearized weak-field regime in the presence of effective de Sitter-like curvature contributions generated by a generic scalar-field potential. It is emphasized that nonlinear screening mechanisms, such as the Vainshtein effect, are not incorporated, as the analysis is restricted to the linear approximation.
The weak-field solutions for a point mass are obtained first in a convenient gauge and subsequently expressed in isotropic and Schwarzschild-type coordinates. For suitable choices of the Horndeski functions, these solutions reduce to well-known forms within specific subclasses of the theory. While Solar-System tests have traditionally been formulated in asymptotically flat spacetimes, the observed accelerated expansion of the Universe motivates considering weak-field configurations containing effective de Sitter-like curvature contributions. Based on the derived metrics, weak-field expressions are obtained for classical Solar-System observables, including perihelion advance, light deflection, and gravitational redshift. Within the weak-field approximation, the influence of the source mass, the minimum of the scalar potential, and the scalar field mass on these observables is systematically examined at linear order. Across all three classical tests, a consistent parameter-dependent structure emerges, primarily controlled by the effective coupling coefficient $\zeta$, which encodes combinations of the background values of $K_{,X}$, $G_{4}(0)$ and $G_{4,\phi}(0)$, and the braiding contribution $G_{3,\phi}(0)$. The scalar-induced corrections to the metric perturbations thus reflect the combined influence of the source mass, the minimum of the scalar potential K(0), and the coupling structure of the underlying Horndeski functions.

The Solar-System tests analyzed here indicate that linearized Horndeski theory exhibits two distinct regimes depending on the scalar-field mass. Analytic solutions are obtained only in the limiting cases of a very light or a very heavy scalar field. In the very light scalar-field regime, Solar-System observations such as those inferred from the Cassini experiment provide a means of assessing the allowed range of the effective coupling parameter $\zeta$
within the linearized weak-field framework adopted here. In this limit, sufficiently large values of $\zeta$ suppress scalar-mediated contributions to perihelion precession, light deflection, and gravitational redshift, so that the resulting predictions approach those of GR at Solar-System scales. At the level of current experimental precision, the corresponding scalar corrections are therefore expected to remain small in local weak-field tests. In the heavy scalar field limit, Yukawa suppression confines scalar-mediated interactions to short ranges, thereby reducing the sensitivity of Solar-System measurements to the parameter  $\zeta$. Although the resulting weak-field predictions closely resemble those of General Relativity locally, background dependent geometric terms associated with the scalar potential K(0) persist in the linearized expansion.
Taken together, these results indicate that the cosmological constant in General Relativity does not generate appreciable corrections at Solar-System scales, in agreement with our previous findings in the BD$\Lambda$ and BDV frameworks. In linearized Horndeski gravity, however, the weak-field structure acquires an explicitly parameter-dependent character, governed by the scalar-field mass and the effective coupling coefficients. While both the very light and very heavy scalar-field limits yield local predictions that closely approximate those of GR, the theory retains residual geometric contributions linked to the minimum of the scalar potential K(0) within the linearized expansion.

The results presented here may motivate further investigation. While the present analysis has been restricted to the linearized regime, future work may examine the role of nonlinear screening mechanisms and their implications for Solar-System observables. Such extensions would allow for a more complete assessment of how local gravitational tests constrain the parameter space of Horndeski models. More broadly, this line of research can deepen our understanding of the role of scalar-field dynamics in local gravitational phenomena.
\section*{Supplementary information}
There is no supplementary file regarding this paper.
\section*{Data availability Statement}
No data associated in the manuscript.

\bibliography{References}

\end{document}